\documentclass[11pt]{article}
\usepackage{amsmath}
\usepackage{amsthm}
\usepackage{graphicx}
\usepackage{multirow}
\usepackage[authoryear]{natbib}
\usepackage{amssymb}
\usepackage{rotating}

\begin{document}
\hspace{13.9cm}


{\LARGE Chaotic Neuronal Oscillations in Spontaneous Cortical-Subcortical Networks}
\ \\
{\bf Pengsheng Zheng}\\
{\small pengsheng.zheng@tuebingen.mpg.de}\\
{\small zheng@fias.uni-frankfurt.de}\\

%


%
\ \vspace{-0mm}\\
\begin{abstract}
Oscillatory activities are widely observed in specific frequency bands of recorded field potentials in different brain regions, and play critical roles in processing neural information. Understanding the structure of these oscillatory activities is essential for understanding the brain function. So far many details remain elusive about their rhythmic structures and how these oscillations are generated. We show that many oscillatory activities in spontaneous cortical-subcortical networks, such as delta, spindle, gamma, high-gamma and sharp wave ripple bands in different brain regions, are genuine chaotic time series which can be reconstructed as chaotic attractors through appropriately selected embedding delay and dimension. The reconstructed attractors are approximated by a simple radial basis function enabling high precision short-term prediction. Simultaneously recorded oscillatory activities in multiple brain regions differ greatly in term of temporal phase and amplitude but can be approximated by the same function. Our results suggest that neural oscillations are produced by deterministic chaotic systems. The occurrence of neural oscillation events is predetermined, and the brain possibly knows when and where the information will be processed and transferred in the future time as a result of the deterministic dynamic.
\end{abstract}

\section{Introduction}
Our brains are foretell machines with advanced predictive powers which are suggested to emerge from various neuronal oscillations~\citep{Buzsaki2006}. These neuronal oscillatory activities prevail in cortical filed potentials in different behavior contexts demonstrating a number of characteristic oscillatory frequency bands, which are crucial for effective information processing within and across distinct brain structures~\citep{Buzsaki2006,Engel2001,Singer1999}. Neural oscillations and synchronization have been linked to many cognitive functions such as sleep and consciousness, perception, motor coordination and memory~\citep{Fell2011,Schnitzler2005,Fries2005}.

Recent findings indicated that network oscillations temporally link neurons into assemblies and facilitate synaptic plasticity~\citep{Buzsaki2004}. The gamma cycle has been suggested to serve as a computational mechanism for the implementation of a temporal coding scheme that enabled fast processing and flexible routing of activity, supporting selection and binding of distributed responses~\citep{Fries2007}. Power-frequency and spatial coherence analyses showed that gamma oscillations in hippocampal CA1 area split into distinct fast and slow frequency components routing the flow of information~\citep{Colgin2009}. Spontaneously occurring oscillatory events like thalamo-cortical sleep spindles and hippocampal sharp wave ripple complexes (SPW-R, 140-250Hz) during slow wave sleep (SWS) were crucial for hippocampal-cortical interactions during memory consolidation~\citep{Wilson1994,Siapas1998,Sirota2003,Diba2007,Eschenko2008,Girardeau2009,Logothetis2012}. Understanding how these oscillations are generated and their rhythmic structures are fundamental and critical for the understanding of brain function.

In this paper, we use two open access data sets, in one of which rats were implanted with silicon probes to monitor
both LFP and unit firing in different hippocampal regions, such as CA1, entorhinal cortex layer 2 (EC2) and layer 5 (EC5) while they were sleeping~\citep{Mizuseki2014}. In the other data set, the recordings were made from anterior thalamus (AT), hippocampus CA1 and mPFC of sleeping mice~\citep{Peyrache2015}.
The simultaneously recorded field potentials in different regions teem with complex rhythmic structures which differ widely in term of phase, amplitude and temporal resolution. However, we show that these diverse rhythmic structures come from deterministic systems, and there is no random components in it.

\section{Reconstruction of neuronal oscillatory activity}
\begin{figure}
\centering
\includegraphics[width=16cm,height=9cm]{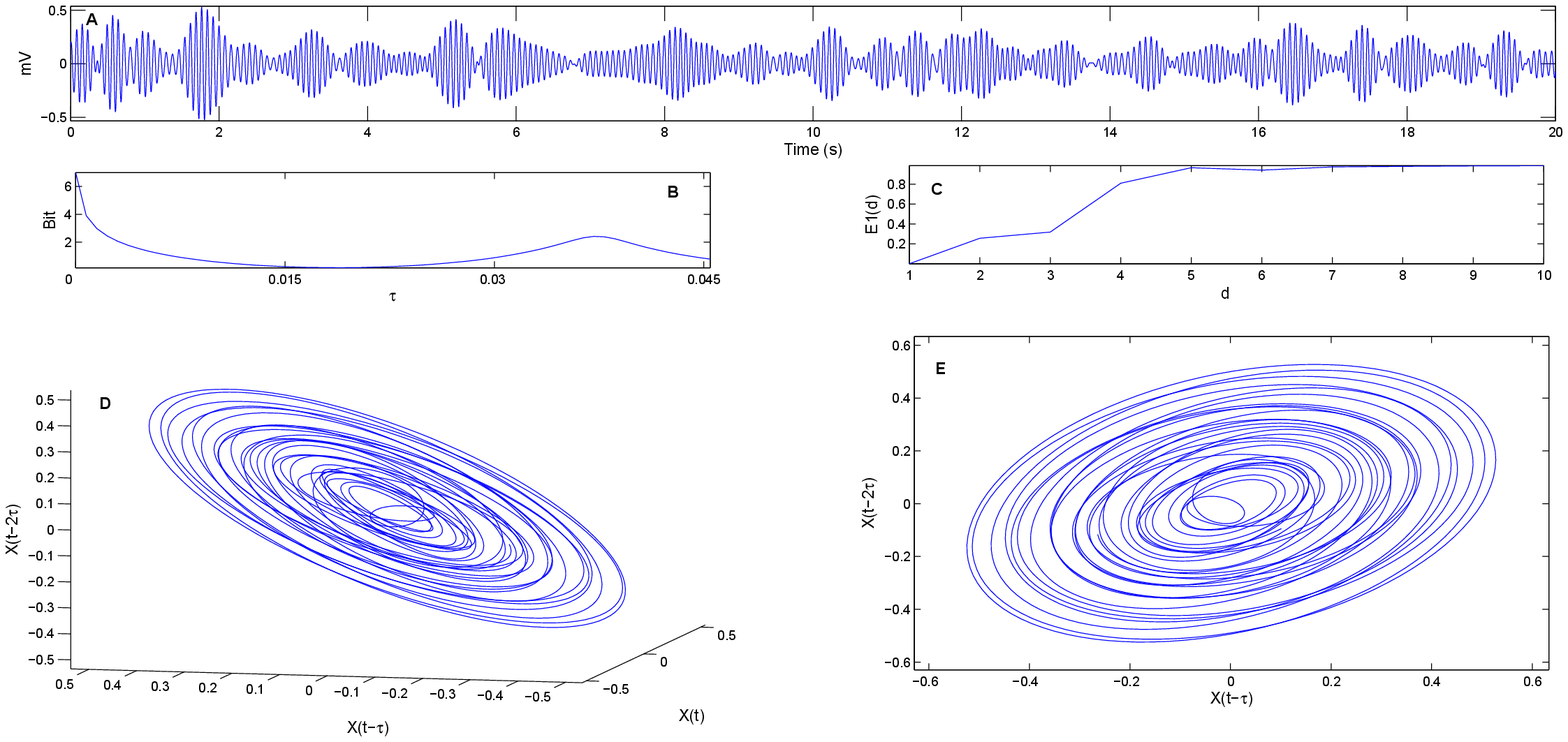}
\caption{Reconstruction of spindle band activity. A: Time series of CA1 spindle.  B: Average mutual information in dependence on embedding delay ($\tau$). C: $E1(d)$, the Cao's-variation from $d$ to $d+1$, as a function of embedding dimension. D, E: Reconstructed attractor projected to 3-D and 2-D spaces respectively.} \label{f1}
\end{figure}

We firstly band-pass filter the LFP to certain frequency range. Unless otherwise stated, we use rat CA1 spindle (12-15Hz) band activity during SWS (ec014.716.dat~\citep{Mizuseki2014}) as an example, and the results were extended to many other oscillatory activities (see Fig.S1-S12 in supplementary material (SM)). The spindle band activity gives us the time sequence of the form $[X(t_0),X(t_1),\cdots,X(t)]$, where $X(t)$ denotes the oscillation voltage at time $t$ (Fig.\ref{f1}A). Following the delay coordinate embedding method~\citep{Takens1981}, the reconstructed attractor of the original system is given by the vector sequence
\begin{equation}
P(t)=[X(t),X(t-\tau),X(t-2\tau),\cdots,X(t-(d-1)\tau)],
\label{e1}
\end{equation}
where $\tau$ and $d$ are the embedding delay and embedding dimension respectively. The delay embedding theorem states that this procedure provides a one-to-one image of the original system for a large enough $d$~\citep{Takens1981}.

The embedding delay ($\tau$) is estimated by mutual information method which measures how much one knows about $X(t-\tau)$ from $X(t)$~\citep{Fraser1986}. In principal, $\tau$ has to be large enough so that the information we get from measuring $X(t-\tau)$ is significantly different from that of $X(t)$, but $\tau$ should not be larger than the typical time in which the system loses information of its initial state.  If $\tau$ is chosen larger than it should, then the reconstructed attractor will be folded and points will look more or less random since they will be uncorrelated. In practice, the $\tau$ can be estimated as the point where the average mutual information (in bits) is close to but before its first local minimum ($\tau=0.015$ in Fig.\ref{f1}B).

The embedding dimension ($d$) is the minimum dimension of the space in which the trajectory does not statistically cross itself. As shown in Fig.\ref{f1}C, the embedding dimension is $d=5$ suggested by Cao's method~\citep{Cao1997}. We then reconstruct the 5-dimensional attractor and project it to 3-D (Fig.\ref{f1}D) and 2-D (Fig.\ref{f1}E) spaces, which are reminiscent of a chaotic attractor (compare Lorenz attractor Fig.S19 in SM).

After reconstructing the attractor, we perform the determinism test to verify if the studied time series are indeed from a deterministic system~\citep{Kaplan1992}. The determinism test constructs the vector field of the system directly from the time series, and subsequently tests if the reconstructed vector field assures uniqueness of solutions in the phase space~\citep{Kaplan1992}, which enables us to distinguish between deterministic chaos and irregular random behavior that often resembles chaos. The calculated determinism measure for reconstructed spindle attractor is 0.99 (1 for perfectly deterministic), which means the system is deterministic.

\section{Neuronal oscillatory activity approximation and prediction}
Following the line of traditional method for chaotic time series prediction, thousands of input $P(t)$ and output $X(t+\tau)$ pairs are collected as $t$ increases from time $0$ to $t_p$, and $t_p$ represents the ending time of training time series. It then reduces to a function approximation problem. Since the system is deterministic, it should be able to find a function to approximate all the input-output pairs. We here use Radial Basis Function Network (RBF) to approximate the delicate input-output function~\citep{Park1991}. The RBF network has simple one-hidden layer structure and fast training process, and most importantly it has a more intuitive mathematical interpretation than the multilayer perceptron as follows,
\begin{equation}
X(t+\tau)=LW\cdot\mbox{Radbas}({\parallel}IW-P(t){\parallel}\cdot{b_1})+b_2,
\label{e2}
\end{equation}
where $IW$ is input weight matrix, $LW$ is weight matrix of output layer, $b_1$ and $b_2$ are bias terms of input and output layers respectively, ${\parallel}\cdot{\parallel}$ represents the Euclidean distance, and
$$\mbox{Radbas}(x)=e^{x^2}.$$
Here we use `newrb' function in the Matlab neural network toolbox~\citep{Beale2013} to find proper parameters $IW$, $LW$, $b_1$ and $b_2$ for equation (\ref{e2}).
The RBF network reaches a small goal of training error ($\mbox{MSE}=1\mbox{e}-6$) in a few iterations implying all the input-output pairs are well-fitted by the same network. The weight matrixes and bias terms of RBF network are frozen after training.

\begin{figure}
\centering
\includegraphics[width=15cm,height=9cm]{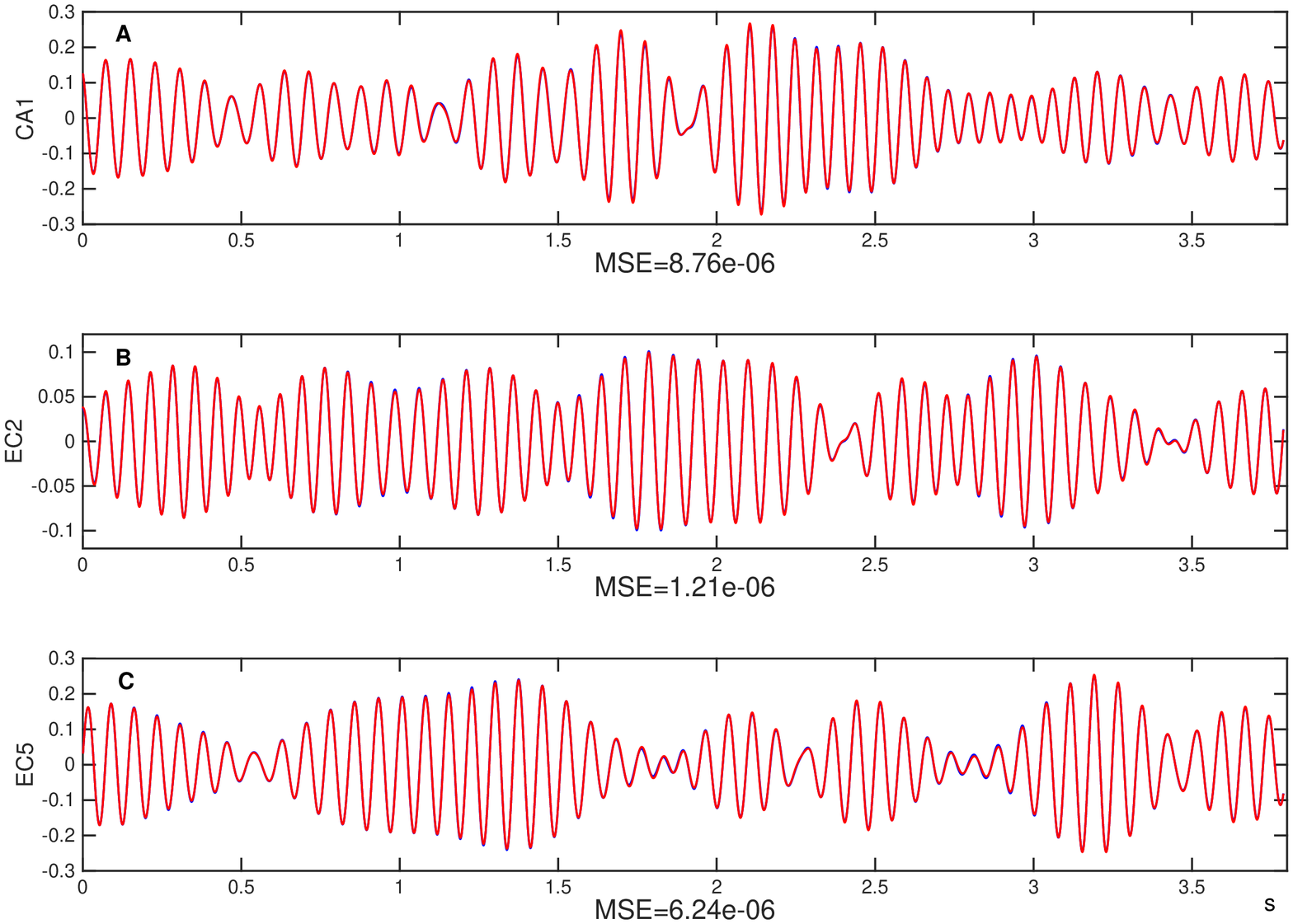}
\caption{Short-term prediction of spindle band activity. A-C, Comparison of predicted (red) and actual (blue) signals in CA1, EC2 and EC5 regions respectively. Note that all the predictions are performed by the network designed for CA1 spindle. MSE is the mean squared error.} \label{f2}
\end{figure}

The prediction is applicable in real time, we are now able to predict the future value $X(t+\tau)$, $t>t_p$, by using $P(t)$ and the frozen equation (\ref{e2}). For short-term prediction, $P(t)$ is always from real signal, but none of these data is exposed to the training of RBF network. Figure \ref{f2}A shows an example of the predicted $X(t+\tau)$, the real signal (blue curve) disappears under the predicted one (red curve) as a result of high-accuracy prediction ($\mbox{MSE}=1.95\mbox{e}-8$). Many other spindle band activities collected from the same rat at different time were also tested. The network learned the data structure from a piece of signal and generalized to unknown signals.  In other words, the spindle band oscillatory activity, no matter in the past or future, can be fitted into the same network (i.e. frozen equation (\ref{e2})), which again proves the spindle activity is dominated by a deterministic system, and there is no irregular or random components in it.

 With the aforementioned method, we further verify that EC2 and EC5 spindle band activities are all developed by a deterministic chaotic system with the same embedding delay and dimension as that of the CA1 region. More interestingly, the EC2 and EC5 spindle band activities are accurately predicted by the RBF network intently designed for CA1 spindle time series (Fig.\ref{f2}B-C). Similarly, an RBF network trained by EC spindle band time series is capable of predicting CA1 spindle band activity. These all suggest that spindle band activities in different regions are dominated by deterministic dynamical systems which share similar or even identical structure.

How could we reconcile the tremendous pointwise differences of simultaneously recorded multi-region signals with the facts that they were approximated by the same function? Take two identical Lorenz systems for example, the difference in initial states leads to vastly different trajectories (See Fig.S20 in SM). This implies the dynamical systems in different regions might be identical in structure but operated with different initial states.

Same argument, Gamma band activities in these three regions have the same embedding delay and dimension, and can be approximated by the same network intently designed for one region. The results can also extend to Delta, High-Gamma and SPW-R band activities with different dynamical systems developed for different frequency bands (see Fig.S1-S12 in the SM).

\begin{figure}
\centering
\includegraphics[width=8cm,height=14cm,angle =-90]{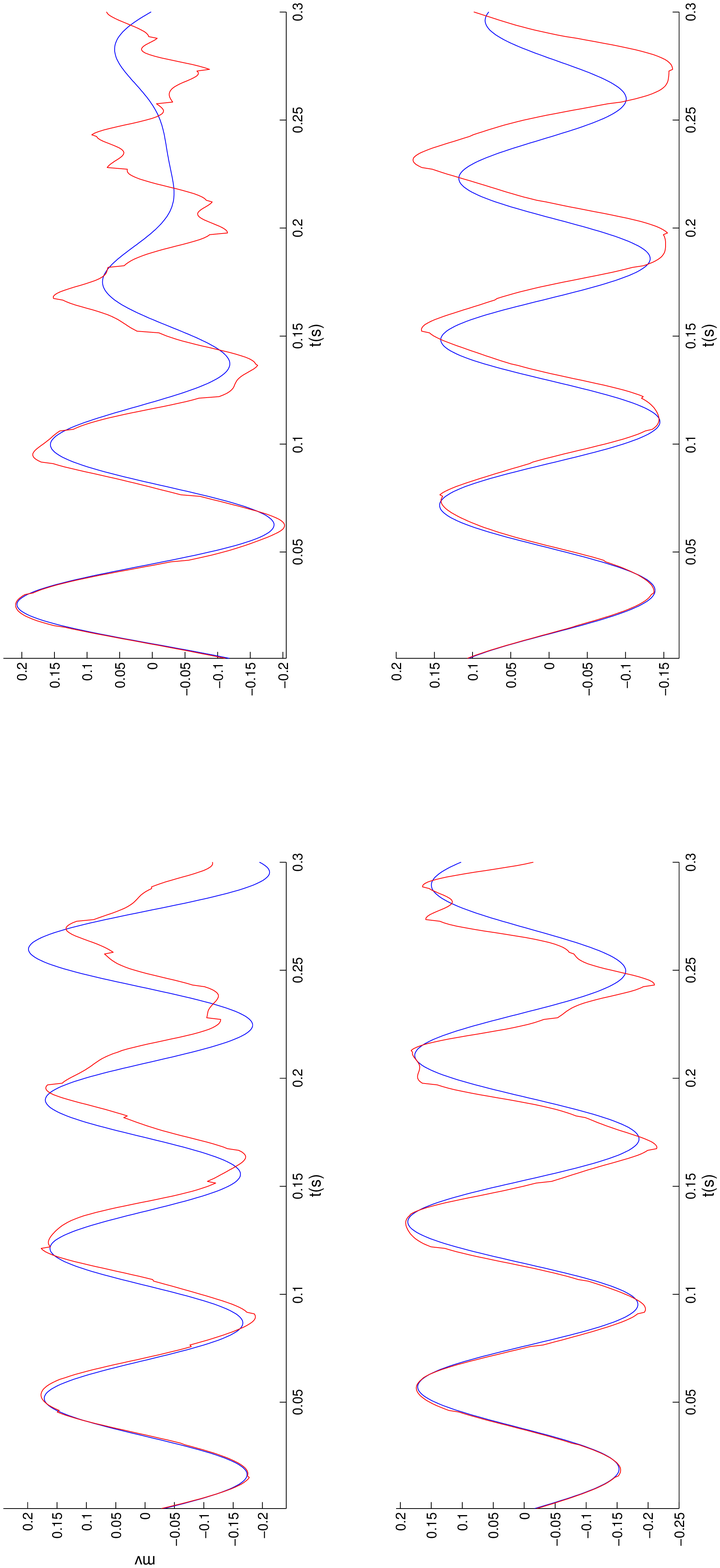}
\caption{Examples of long-term iterative prediction of CA1 spindle activity. Red and blue curves represent the predicted and real signals respectively. Note that none of these data is exposed to the training of the network.} \label{f3}
\end{figure}

In long-term prediction, the predicted value $X(t+\tau)$ can be further applied to the prediction of $X(t+2\tau)$, and so forth. Unlike the short-term prediction, $P(t)$ $(t>t_p+\tau)$ now is composed of predicted values rather than real signals. Figure \ref{f3} shows four examples of long-term iterative prediction of CA1 spindle band activity by the same frozen network. In all cases, the network achieves high-precision prediction at the beginning, and the performances extend for a certain amount of time followed by decreasing precision as time elapses. This is consistent with the observation of typical chaotic systems in which long term prediction is only theoretically plausible within a limited period of time due to its exponential divergence nature.

\begin{figure}
\centering
\includegraphics[width=15cm,height=7cm]{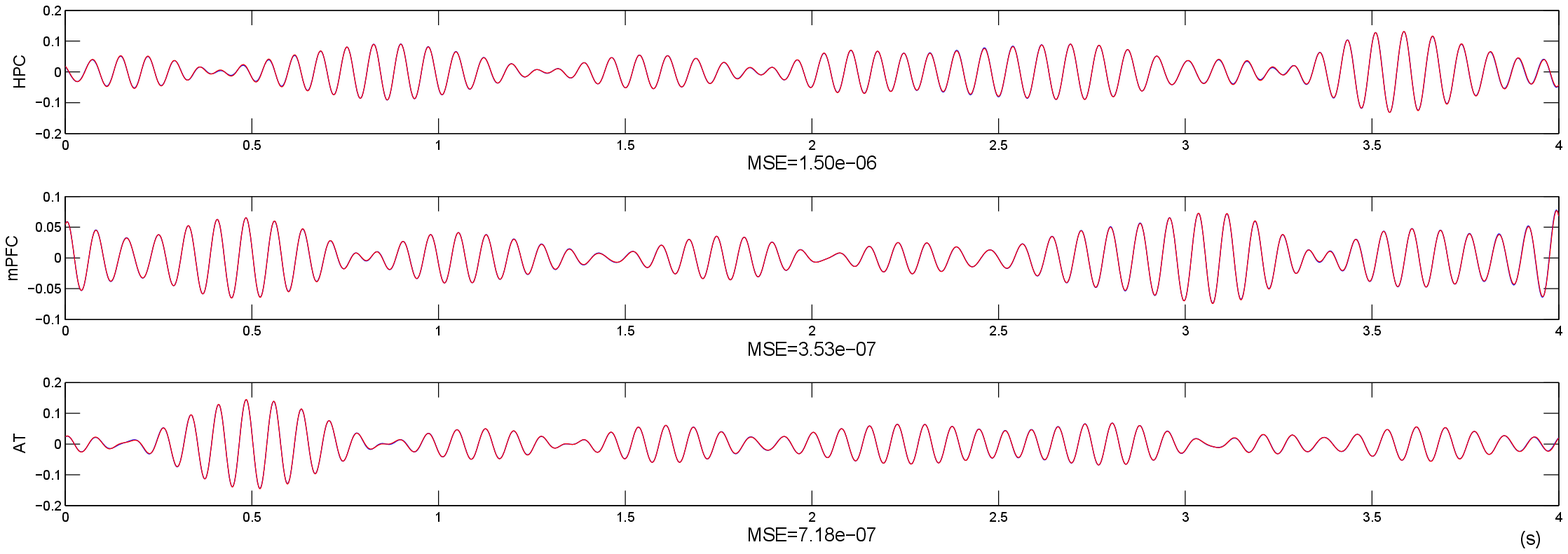}
\caption{Short-term prediction of spindle band activity in sleeping mouse (Mouse12-120807). Comparison of predicted (red) and actual (blue) time series in hippocampal (HPC) CA1, mPFC and AT regions respectively. Note that all the predictions are performed by the network designed for HPC CA1 spindle.} \label{f4}
\end{figure}

All the results are further verified on the open access data set recorded in Hippocampal-mPFC-thalamic network of sleeping mice~\citep{Peyrache2015}, an RBF network is designed with the time series from one brain region and then applied to the prediction of signals in other brain regions. Figure \ref{f4} shows the results of short-term prediction of spindle band activity. More technical details and the results of other frequency bands are reported in Fig.S16-S17 (SM). Unless otherwise stated, all the results are retrieved during SWS state, but the network designed during SWS state is capable of predicting the signal in REM state (Fig.S18 SM). Matlab code is provided in the SM for easy and fast test of users'' own neural signal under different conditions.

\section{Conclusion and discussion}
Many cognitive functions of the brain are suggested to be emerged from neural oscillations, but where do these complex oscillations come from?
Here we show that these diverse rhythmic activities, in different brain regions but in the same frequency band, come from a unified deterministic system, and there is no random components in it. The occurrence of certain kinds of neural oscillation events is predetermined by the deterministic dynamic rather than randomly or statistically formed. In this sense, the brain has perfect control of local and global information flow, which is ideal for information transfer, avoidance and collaboration between distinct regions.

There is a long history of chaotic system study in physics and many interdisciplinary areas with numerous proposed theorems and applications. Now the whole theory is potentially open to neural oscillation research, and many interdisciplinary topics are expected to be explored in the future. On the other hand, neural oscillations have been suggested to play critical roles in many cognitive functions such as sleep, consciousness, perception and memory. The results of paper provide potential new perspectives to many of these researches. For example,
among all the topics of chaos theory, the synchronization of multiple chaotic attractors is one of the hottest topics. As mentioned above, two identical Lorenz systems display vastly different trajectories when they are operated at different initial states (Fig.S21 in SM). However, the two systems can be fast synchronized if a negative feedback is applied (Fig.S15 in SM). The synchronization level is further adjustable by tuning the strength of the negative feedback which might be analogue to the changes of network connectivity wildly observed in epilepsy~\citep{David2008,Zhang2011,Clemens2013}. Integration of the results with the epilepsy research might be one of the interesting studies in the future.

\bibliographystyle{apa}
\bibliography{ref}

\end{document}